\documentclass[12pt,preprint]{aastex}
\usepackage{amsmath}
\usepackage{graphicx}
\usepackage{natbib}

\begin{document}
\newcommand{\skipthis}[1]{}
\newcommand{\hii}{{\rm H}{\sc ii}}
\newcommand{\uchii}{{\rm UCH}{\sc ii}}
\def\um{$\mu$m}
\def\nh2d{$\rm{NH_2D}$}
\def\nh3{$\rm{NH_3}$}
\def\NH3{$\rm{NH_3}$}
\def\n2hp{$\rm{N_2H^+}$}
\def\NTH{$\rm{N_2H^+}$}
\def\msolar{$M_\odot$}
\def\msun{$M_\odot$}
\def\Mout{$\dot{M}_{\rm out}$}
\def\Macc{$\dot{M}_{\rm acc}$}
\def\lsolar{$L_\odot$}
\def\lsun{$L_\odot$}
\def\kms-1{km~s$^{-1}$}
\def\h2o{$\rm{H_2O}$}
\def\h2{$\rm{H_2}$}
\def\hcop{$\rm{HCO^+}$}
\def\cm2{$\rm{cm^{-2}}$}
\def\cm3{$\rm{cm^{-3}}$}
\def\vlsr{$\rm{V_{LSR}}$}
\def\vmax{$v_{\rm max}$}
\def\tdyn{$t_{\rm dyn}$}
\def\tff{$t_{\rm ff}$}
\def\uv{$(u,v)$}
\newcommand{\lsim}{${\raisebox{-.9ex}{$\stackrel{\textstyle<}{\sim}$}}$ }
\newcommand{\gsim}{${\raisebox{-.9ex}{$\stackrel{\textstyle>}{\sim}$}}$ }

\slugcomment{Accepted to ApJ}

\shorttitle{Hierarchical Fragmentation and Jet-like Outflows in IRDC G28.34+0.06}
\shortauthors{Ke Wang et al.}

\title{Hierarchical Fragmentation and Jet-like Outflows in IRDC G28.34+0.06, a Growing Massive Protostar Cluster}

\author{
Ke Wang\altaffilmark{1,2},
Qizhou Zhang\altaffilmark{2},
Yuefang Wu\altaffilmark{1},
and Huawei Zhang\altaffilmark{1}
}

\email{kwang@cfa.harvard.edu}
\email{qzhang@cfa.harvard.edu}

\altaffiltext{1}{Department of Astronomy, School of Physics, Peking University,
Beijing 100871, China}
\altaffiltext{2}{Harvard-Smithsonian Center for Astrophysics, 60 Garden Street,
Cambridge MA 02138, USA}

\begin{abstract}
We present Submillimeter Array (SMA) $\lambda = 0.88$\,mm observations 
of an infrared dark cloud G28.34+0.06.
Located in the quiescent southern part of the G28.34 cloud,
the region of interest is a massive ($>10^3$\,\msun) molecular clump P1 
with a luminosity of $\sim 10^3$ \lsun, 
where our previous SMA observations at 1.3\,mm have revealed 
a string of five dust cores of 22--64 \msun\ along the 1 pc IR-dark filament. 
The cores are well aligned at a position angle of 48$^{\circ}$
and regularly spaced at an average projected separation of 0.16 pc. 
The new high-resolution, high-sensitivity 0.88\,mm image further resolves 
the five cores into ten compact condensations of 1.4--10.6 \msun, 
with sizes a few thousands AU.
The spatial structure at clump ($\sim 1$ pc) and core ($\sim 0.1$ pc) scales 
indicates a hierarchical fragmentation. 
While the clump fragmentation is consistent with a cylindrical collapse, 
the observed fragment masses are much larger than the expected thermal Jeans masses. 
All the cores are driving CO\,(3--2) outflows up to 38 \kms-1, 
majority of which are bipolar, jet-like outflows. 
The moderate luminosity of the P1 clump sets a limit on the mass of protostars of 3--7 \msun.
Because of the large reservoir of dense molecular gas in the immediate medium 
and ongoing accretion as evident by the jet-like outflows, 
we speculate that P1 will grow and eventually form a massive star cluster. 
This study provides a first glimpse of massive, 
clustered star formation that currently undergoes through an intermediate-mass stage.
\end{abstract}

\keywords{
ISM: individual (G28.34+0.06) --- ISM: jets and outflows
--- stars: formation --- stars: early-type
}

\section{Introduction} \label{sect:intro}
Stars are born in dense molecular clouds through gravitational collapse.
A classic picture of low-mass star formation has been described in \cite{shu87}.
During the collapse, a parsec-scale molecular clump fragments into gaseous
cores that subsequently collapse and fragment to condensations which eventually
form individual or a group of stars. 
After the formation of an accretion disk,
a bipolar wind develops to shed angular momentum in the disk so that circumstellar
matter can accrete onto the protostar.
Interactions between such a wind and the ambient molecular cloud may produce highly collimated
molecular outflows (or ``molecular jets'') at parsec scales \citep{shang2006, shang2007}.

Massive protostars are often found embedded in massive molecular clumps of $10^3$ \msun\
\citep{plume97, molinari2000, beuther02, beltran06}.
Their presence is marked by high luminosity ($> 10^3$ \lsun), 
copious spectra from organic molecules \citep{blake87,Schilke1997}, 
active accretion \citep{zhang2005a,Cesaroni2006}, and
strong molecular outflows \citep{zhang2001,zhang2005,Beuther02c}. 
However, the stage prior to the high mass protostellar phase is
not clear. Do massive protostars evolve through a low to intermediate-mass stage? How does
accretion proceed? Do cores harboring massive young stellar objects collect all
the mass before the formation of a protostar? These questions can potentially be addressed
by studying massive infrared dark clouds (IRDCs).

IRDCs are dense ($>10^4$\,\cm3) and cold ($<25$\,K) molecular clouds that
absorb Galactic background mid-infrared emission while emit at longer wavelengths ($\gtrsim$ 100\,\um).
Because of their extreme properties,
IRDCs are ideal laboratories to study the early stages of star formation,
and thus have been subjected to intense studies since their identification
in mid-1990s by the \emph{Infrared Space Observatory (ISO)} and the \emph{Midcourse Space Experiment}
\citep[\emph{MSX};][]{perault96,egan98,hennebelle01}.
Comprehensive catalogs have been established based on Galactic IR surveys from the
\emph{MSX} \citep{simon06a, simon06b} and \emph{Spitzer} \citep{pretto10}, 
revealing more than 10,000 IRDCs.
Among them, the ones with masses of $10^3$ \msun\ and luminosities $< 10^3$ \lsun\
are candidates of massive star formation at a stage prior to the high mass protostellar phase.

G28.34+0.06 (hereafter G28.34), 
one of the first recognized and well studied IRDCs \citep{carey98, carey2000},
appears to be a candidate to form massive stars.
The giant cloud, at a distance of $\sim$4.8\,kpc, contains several $10^3$\,\msun\ of 
dense gas traced by 1\,mm dust emission along the IR-dark filaments
extending $\sim$6\,pc \citep{carey2000,simon06b,rath06,pillai06,wang08,zhang09}.
Two prominent dust clumps, P1 and P2,
are revealed from dust continuum images
obtained from single dish telescopes \citep{carey2000, rath06}.
Large amount of dense gas ($\sim 10^3$\,\msun\ within $<1$\,pc) makes
P1 and P2 potential sites of protostellar cluster formation.
\cite{zhang09} observed P1 and P2 with the SMA at 230\,GHz band.
They found that, despite similar amount of dense gas contained in the two clumps,
P1 and P2 show very different stages of evolution.
At a $\sim 1''$ resolution,
the northern clump P2 fragments to two compact cores with masses 49 and 97\,\msun, respectively.
The cores emit rich organic molecular line emissions analogous to that observed in hot cores.
The southern clump P1, on the other hand,
fragments to a string of five cores regularly spaced along the IR-dark ridge,
with masses of 22--64\,\msun.
In contrast to the cores in P2,
the P1 cores show no molecular line emission except faint CO(2--1)
across the entire 4\,GHz SMA band.
This implies a significant CO depletion at a scale over 0.07\,pc in the P1 cores.
In addition, an enhancement of deuterium fractionation was also found toward P1 \citep{chen10}.
All these properties point P1 to a quiescent stage much earlier than P2.
However, strong H$_2$O maser \citep{wang06} and enhanced 4.5\,\um\ emission
\citep[``green fuzzy'';][]{chambers09},
as well as a faint 24\,\um\ source in P1 suggest that star formation may be ongoing.

Here we report our new SMA observations of G28.34-P1 at the 345\,GHz band.
The high angular resolution, high sensitivity observations 
reveal hierarchical structures and multiple jet-like outflows,
which suggest that intense star formation is already ongoing in this region.

\section{Observations and Data Reduction} \label{sect:obs}

We observed the molecular clump P1 in IRDC G28.34+0.06 with the Submillimeter Array
\footnote{The Submillimeter Array is a joint project between the Smithsonian Astrophysical Observatory
and the Academia Sinica Institute of Astronomy and Astrophysics and is funded by the
Smithsonian Institution and the Academia Sinica.}
\citep[SMA;][]{Ho04} at the 345\,GHz band from 2009 April 26 through 2009 September 23.
During six observing runs, the SMA antennas were in
three different array configurations: compact (COM), extended (EXT), and subcompact (SUB),
each was used to observe for two nights for this project.
Table \ref{tab:obs} summarizes the observations.

For all the observations, we used a common phase center \\
(R.A., decl.)
$_{\mathrm J2000}=(18^{\mathrm h}42^{\mathrm m}50.\!^{\mathrm s}74, -04^{\circ}03^{'}15.\!^{''}3)$.
Quasars J1751+096 and J1830+063 were observed approximately every 20 minutes to monitor time dependent antenna gains.
The receivers were tuned to a local oscillate frequency of 341.6\,GHz,
with a uniform spectral resolution of 0.812\,MHz (or 0.7\,\kms-1) across the entire band.
The system temperature varies from 120 to 500\,K during the six tracks.
The full-width-half-maximum (FWHM) primary beam is about 34$''$.

For the COM observations, we used quasar 3C273 for frequency dependent bandpass calibration.
Absolute flux was obtained by comparing the observed correlator counts with modeled fluxes of dwarf planet Ceres.
With an IF of 4--6\,GHz, the lower sideband (LSB) covered rest frequencies from 335.7 
through 337.7\,GHz and the upper sideband (USB) covered 345.6--347.6\,GHz.

For the EXT observations, 3C454.3 was used for bandpass calibration,
while young star MWC349 or planet Uranus was observed for flux calibration.
The observations made use of the newly upgraded capability that
doubled the original IF bandwidth to 4\,GHz in each sideband, covering
rest frequencies 333.7--337.7\,GHz in the LSB and 345.6--349.6\,GHz in the USB.

For the SUB observations, bandpass was calibrated by observing 3C454.3
and flux was scaled by observing Jovial satellites Callisto and Ganymede.
The frequency coverage was the same as the EXT observations.

The visibility data were calibrated using MIR IDL software package.\footnote{\url{ http://www.cfa.harvard.edu/$\sim$cqi/mircook.html}}
Calibrated visibility data were then exported to MIRIAD\footnote{\url{ http://www.cfa.harvard.edu/sma/miriad}} for further processing and imaging.
CASA\footnote{\url{ http://casa.nrao.edu}} was also used for part of the image processing.

Data from different observing runs were calibrated separately,
and then combined in the visibility domain for analysis.
Continuum emission was generated by averaging line free channels in the visibility domain.
In this study, 
if not otherwise stated,
we use the EXT data for continuum analysis
and the combined SUB and COM data for spectral analysis.
With natural weighting, 
the synthesized beam is $0''.69\times 0''.64$, with $\rm{PA}=-83^{\circ}.1$ for the continuum image;
while for the spectral line, the beam is $3''.23\times 1''.96$, $\rm{PA}=-47^{\circ}.4$.
The $1\sigma$ rms noise is about 0.8\,mJy in the continuum, 
and about 100\,mJy (0.2\,K) per 1\,\kms-1 in the spectra.
If we use the SUB data only, 
the beam becomes larger and we obtain a $1\sigma$ rms of 0.1\,K at the same spectral resolution.
Empirically, the absolute flux is accurate to about 15\%.
The uncertainty in the absolute position is $\lesssim 0''.1$,
obtained by comparing the fitted position of the secondary gain calibrator and its catalog position.

\section{Results and Discussion}
\subsection{Continuum data and fragmentation}
\subsubsection{Hierarchical fragmentation \label{sect:frag}}

There are different definitions of clump, core, and condensation in the literature
when describing the spatial structure of dust continuum or molecular line emission.
In the following discussion,
we refer a {clump} as a structure with a size of $\sim 1$ pc,
a {core} as a structure with a size of $\sim 0.1$ pc,
and a {condensation} as a substructure of $\sim 0.01$ pc within a core.
A clump is capable of forming a cluster of stars,
a core may form one or a small group of stars,
and a condensation can typically form a single star or a multiple-star system.
This nomenclature is consistent with the one adopted by \cite{zhang09}.

\begin{figure}
\includegraphics[width=\textwidth,angle=0]{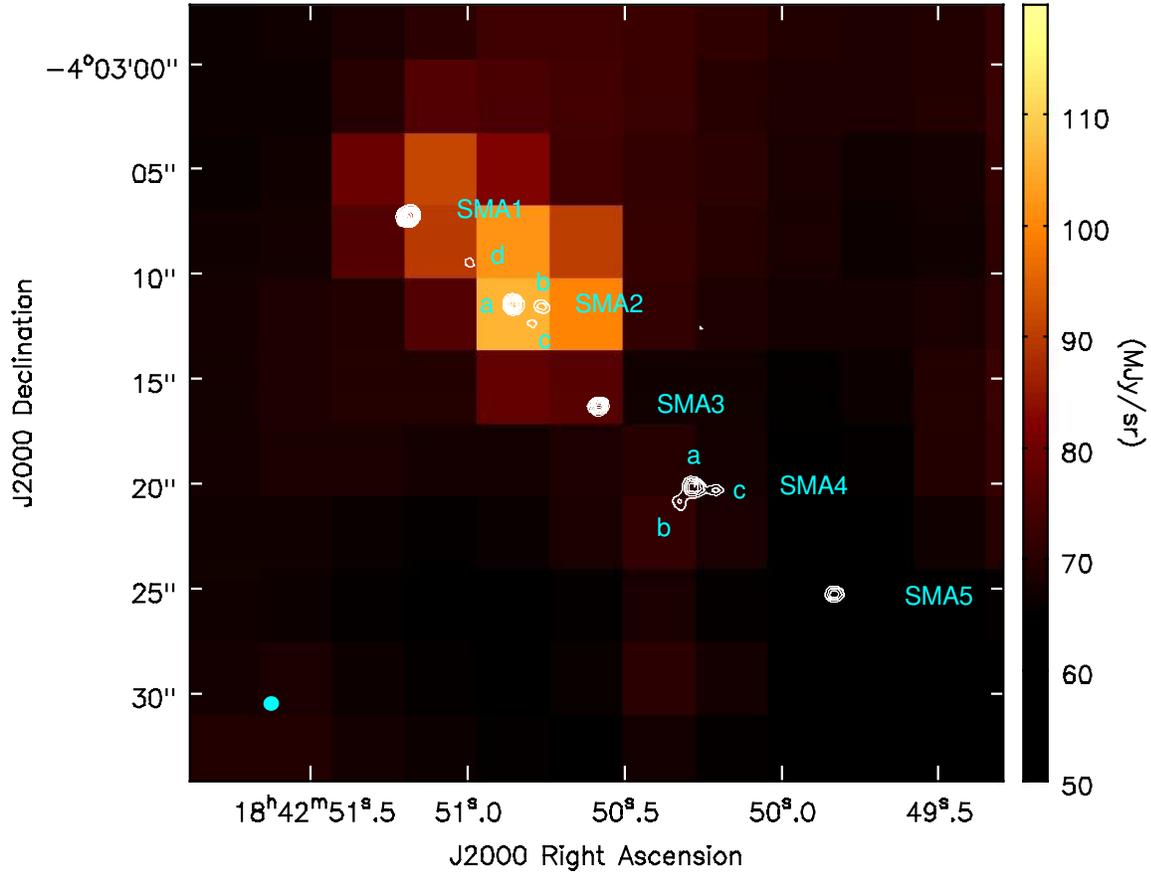}
\caption{SMA 0.88\,mm image of the G28.34-P1 region in contour
overlaid on the \emph{Spitzer} 24\,\um\ image in color scale.
The SMA image is made from the EXT configuration data only,
with synthesized beam $0''.69\times 0''.64$, $\rm{PA}=-83^{\circ}.1$,
indicated as an ellipse 
in the lower left corner of the panel.
The contours start at 4\,mJy ($5\sigma$) and increase by a step of
1.6\,mJy ($2\sigma$).
Assigned condensation names are also labeled on the image.
The SMA image shown here is not corrected for the primary beam attenuation.
\label{fig:cont}}
\end{figure}

Figure \ref{fig:cont} presents the 0.88\,mm continuum emission.
The $0''.69 \times 0''.64$ resolution image resolves the clump 
into five groups of 
compact condensations,
corresponding to the five dust cores discovered at 1.3\,mm,
namely SMA1, SMA2, SMA3, SMA4, and SMA5 \citep{zhang09}.
Cores SMA1, SMA3, and SMA5 consist of one condensation.
Cores SMA2 and SMA4 consist of three condensations, 
with one dominating over the other two in peak flux.
In addition, a faint condensation is also revealed northeast to core SMA2.
We tentatively assign this condensation to the SMA2 group.
We name all the 10 condensations accordingly and label them on Fig. \ref{fig:cont}.
Refining the core positions using the positions of the dominant condensations,
the five cores are spaced by an projected distance of $0.16 \pm 0.02$\,pc ($6''.8 \pm 1''.0$),
and are well aligned at a position angle (PA, east of north) of 48$^{\circ}$.
In cores SMA2 and SMA4, 
the three condensations are spaced by $0.03 \pm 0.007$\,pc ($1''.3 \pm 0''.3$).

We estimate the mass of each condensation.
Assuming optically thin dust emission, the dust mass can be estimated following
$$
M_{\rm dust}=\frac{F_{\nu}d^2}{B_{\nu}(T_{\rm dust}){\kappa}_{\nu}}\,,
$$ 
where $M_{\rm dust}$ is the dust mass, $F_{\nu}$ is the continuum flux at
frequency $\nu$, $d$ is the source distance, $B_{\nu}(T_{\rm dust})$ is the
Planck function at dust temperature $T_{\rm dust}$, and
${\kappa}_{\nu}=10({\nu}/1.2\,{\mathrm {THz}})^{\beta}$
cm$^2$\,g$^{-1}$ is the dust opacity \citep{hildebrand83}.
We adopt the dust temperatures from \cite{zhang09}, {i.e.},
16\,K for SMA2 and 13\,K for other cores.
The dust opacity at millimeter/sub-millimeter wavelengths is uncertain.
For dust in the interstellar medium, $\beta=2$ \citep{Draine1984}.
\cite{rath10} derived $\beta=1.5 \pm 0.3$ for the entire P1 clump based on 
a global spectral energy distribution.
We generated a dust opacity map by comparing the continuum image at 345 GHz and 230 GHz,
and found $\beta$ to vary from 1.3 to 2.5 across the map. Given the uncertainties,
we adopt $\beta=1.5$ for all the cores similar to \cite{zhang09},
and discuss the effect of different $\beta$ on the mass estimates.

Table \ref{tab:cores} lists coordinates, size, peak flux, integrated flux, 
and estimated mass of all the 10 condensations.
These parameters are determined by fitting a two dimensional Gaussian function
to the observed flux distribution.
Fluxes are then corrected for primary beam attenuation.
The masses of the condensations range from 1.4 to 10.6\,\msun\ with an average of 4.8\,\msun.
The mean size is $0''.9 \times 0''.7$, or $0''.6 \times 0''.2$ 
after deconvolved with the synthesized beam.
The deconvolved size corresponds to about $3000 \times 1000$\,AU at the source distance.
The uncertainty in the mass estimation arises from several factors.
First, the flux calibration contributes 15\,\% in uncertainty.
Second, the uncertainty of $\beta$ contributes in a form of
$M_{dust} \propto 3.5^{\beta}$.
Third, the spatial filtering effect of radio interferometry
preferentially picks up the compact structures
and filters out the extended emission.
Comparing images made from different array configurations,
the EXT image recovered 20\,\%--30\,\% fluxes of what covered by COM and SUB images.
This filtering effect explains why the masses reported in Table 2 are consistently 
lower than those in \cite{zhang09}. The EXT image at 345 GHz represents the high 
contrast structure in these cores
--- the small, compact condensations tracing the immediate surroundings of the protostars.

Our observations reveal 
fragmentation at different spatial scales
in G28.34-P1.
First, the 1 pc clump fragments into five cores \citep{zhang09};
second, two of the 0.1 pc cores fragment into even small condensations of 0.01 pc.
The fragmentation at clump and core scales are consistent with
a hierarchical fragmentation picture. 
The clump fragmentation is likely the result of initial physical conditions
(density, temperature, turbulence, and magnetic fields),
while the core fragmentation results from an increased density after the initial fragmentation.
Based on the Very Large Array (VLA) \nh3\ (1,1) and (2,2) spectra \citep{wang08} and 
the IRAM 30m 1.3\,mm continuum image \citep{rath06},
we measure an initial gas temperature of 13\,K and density of $3\times10^5$ cm$^{-3}$,
over a scale of 1 pc toward the P1 clump.
These values yield a thermal Jeans mass of 0.8\,\msun.
At the core scale, density increases to several 10 times of the initial clump density.
For instance, in SMA2 the core density measured from the 230\,GHz image
is $9.6\times10^6$ cm$^{-3}$,
32 times of the initial clump density.
This density, together with an increased gas temperature (16 K),
leads to a thermal Jeans mass of 0.2\,\msun.
These parameters are also listed in Table \ref{tab:frags}.

It is worth stressing that,
in both clump and core fragmentation, the observed fragment (core/condensation) masses
are significantly ($\sim 10$ times or more) larger than the 
expected thermal Jeans masses (Table \ref{tab:frags}),
indicating that thermal pressure is not dominant in the fragmentation processes.
Instead, turbulence may play an important role in supporting
these large core/condensation masses, as discussed in \cite{zhang09}
(see further discussion in Section \ref{subsect:cylinder}).
The 1$\sigma$ rms sensitivity in Fig.~\ref{fig:cont} is 0.8 mJy, corresponding to 0.2\,\msun,
well below the thermal Jeans mass at the clump scale (0.8\,\msun).
However, we did not detect any cores of 0.8\,\msun\ in the filament.
The non-detection of Jeans mass cores in the clump may suggest that
low mass cores are not significantly centrally peaked,
thus they cannot be substantiated from the smooth emission in the clump.

\subsubsection{Cylindrical collapse \label{subsect:cylinder}}

The most intriguing feature in G28.34-P1
is the configuration of well aligned, regularly spaced cores.
The alignment and regularity strongly suggest a mechanism that is responsible for 
shaping the molecular filament into what it is now.
Similar features (but at different spatial scales) have been reported 
in a large number of nearby dark clouds \citep{Schneider1979},
in a few filamentary IRDCs \citep[e.g.,][]{Jackson2010, Miettinen2010}, 
as well as in numerical simulations \citep{Martel2006}.
It has been suggested that these fragments are most likely the results of 
gravitational collapse of a cylinder,
as first proposed by \cite{Chandra1953} and followed up by \cite{Nagasawa1987}
\citep[also see discussion in][]{Jackson2010}.
The theory predicts that, under self gravity,
the gas in the cylinder will ultimately beak up into pieces with a certain interval ($\lambda _{\rm max}$)
at which the instability grows the fastest and thus dominates the fragmentation process.
Such `sausage' instability produces a chain of equally spaced fragments along the filament,
with the spacing roughly the interval $\lambda _{\rm max}$.
The mass per unit length along the cylinder (or linear mass density) has a maximum value,
$(M/l)_{\rm max} = 2v^2/G = 465\,\left(\frac{v}{\rm km\,s^{-1}}\right)^2$ \msun\,pc$^{-1}$, 
where $G$ is the gravitational constant.
If the cylinder is supported by thermal pressure, $v$ is the sound speed $c_s$;
if, on the other hand, it is mainly supported by turbulent pressure, 
then $v$ is replaced by the velocity dispersion $\sigma$.

We now compare the observational results with the theoretical predictions.
In G28.34-P1, the FWHM line width is 1.7 \kms-1, 
measured form the \nh3 (1,1) spectrum \citep{wang08,zhang09}.
The line width is equivalent to a velocity dispersion of $\sigma = 0.72$ \kms-1 
if the line profile is Gaussian.
This velocity dispersion leads to a maximum mass density of 240 \msun\,pc$^{-1}$.
The total core mass in P1 is 183 \msun\ over 0.8 pc along the filament \citep{zhang09},
implying a linear mass density of 230 \msun\,pc$^{-1}$.
Taking into account the missing flux in the interferometer maps,
the observations match the theoretical predictions well 
if G28.34-P1 is mainly supported by turbulence.

The separation between two adjacent fragments, $\lambda _{\rm max}$,
depends on the nature of the cylinder.
For an incompressible fluid cylinder in absence of magnetic field,
$\lambda _{\rm max} = 11R$, where $R$ is the unperturbed cylindrical radius.
And the instability manifests at about two free-fall timescales (see Section \ref{subsect:outflow_para}).
In G28.34-P1, the average separation between cores is $\lambda _{\rm max} = 0.16$\,pc,
implying an initial radius of $R = 0.014$\,pc,
several times smaller than the core diameter of 0.1 pc \citep{zhang09}.
If magnetic field is involved, the strength is of the order of 
$B = 4\pi \rho R\sqrt{G}$ \citep[see Equation (93) in][]{Chandra1953}.
With $n = 3\times 10^5$ \cm3\ and $R=0.014$ pc,
we obtain a magnetic field strength of 0.16 mG.
This is relatively small compared to magnetic field strength observed in 
massive star formation regions \citep[e.g.,][]{girart2009}.

For an isothermal gas cylinder,
$\lambda _{\rm max} = 22H$, where $H = v(4\pi G\rho_c)^{-1/2}$ is the scale height,
whereas $\rho_c$ is the gas density at the center of the cylinder.
The equation can be simplified as
$$
\lambda _{\rm max} = \left\{ \begin{array}{rl}
 0.15\, \rm{pc}\, \left(\frac{c_s}{\rm 0.21\, km\,s^{-1}}\right) 
\left(\frac{n}{3\times 10^5 \,{\rm cm^{-3}}}\right) 
&\mbox{ for thermal support,} \\
 0.13\, \rm{pc}\, \left(\frac{\sigma}{\rm 0.72\, km\,s^{-1}}\right) 
\left(\frac{n}{5\times 10^6 \,{\rm cm^{-3}}}\right) 
&\mbox{ for turbulent support.}
       \end{array} \right.
$$
In the above equations, 
the sound speed $c_s$ of 0.21 \kms-1 is calculated using a gas temperature of 13 K;
volume density $3\times 10^5$ \cm3 is the averaged clump density,
while $5\times 10^6$ \cm3 is the average of 
clump density ($3\times 10^5$ \cm3) and core density ($9.6\times 10^6$ \cm3)
given in Table \ref{tab:frags}.
We see that both scenarios are in agreement with the observed spacing of $\lambda _{\rm max} = 0.16 \pm 0.02$\,pc
if the assumptions are reasonable.
However, since the density should be the central density of the cylinder,
we suspect that the turbulent support is more likely dominant in G28.34-P1.

Comparison between observational results and theoretical predictions suggests that
the fragmentation in the G28.34-P1 filament is well represented by a cylindrical collapse.
The filament is likely supported mainly by turbulence rather than thermal pressure.

\subsubsection{Core structures}

We measure the core density profiles in unresolved cores SMA1, SMA3, and SMA5.
The measurements are performed in the visibility domain to avoid defects in image deconvolution.
We combine the SUB, COM, and EXT data for analysis for this purpose.
The combined visibility data sample the G28.34-P1 region over baseline lengths
ranging from about 5 to 200 k$\lambda$,
offering a fairly well sampled \uv\ measurement.

For a spherical core, the observed flux distribution over radial distance $F(r)$
represents the radial density profile modified with temperature gradient.
The dust temperature scales as $T_{dust} \propto r^{-a}$ \citep{scoville76} 
with $a=0.33$, if the core is internally heated.
As long as Rayleigh-Jeans approximation holds,
this temperature gradient contributes to the total flux in the form of $\propto r^{-a}$.
Assuming a power law density profile, $\rho \propto r^{-b}$, and optically thin dust emission,
the observed dust continuum flux integrated along the line of sight,
$F \propto \int \rho T_{dust} ds$, or
$$F \propto r^{-(a + b - 1)}\,,$$
when $(a+b)>1$.
In the visibility domain, this is Fourier transformed into a form of
$$ A_{uv} \propto S_{uv}^{(a + b - 3)}\,,$$
where $A_{uv}$ is the visibility amplitude
and $S_{uv} = \sqrt{(u^2 + v^2)}$ is the \uv\ distance \citep{looney2000,zhang09}.

\begin{figure}
\includegraphics[width=0.5\textwidth,angle=0]{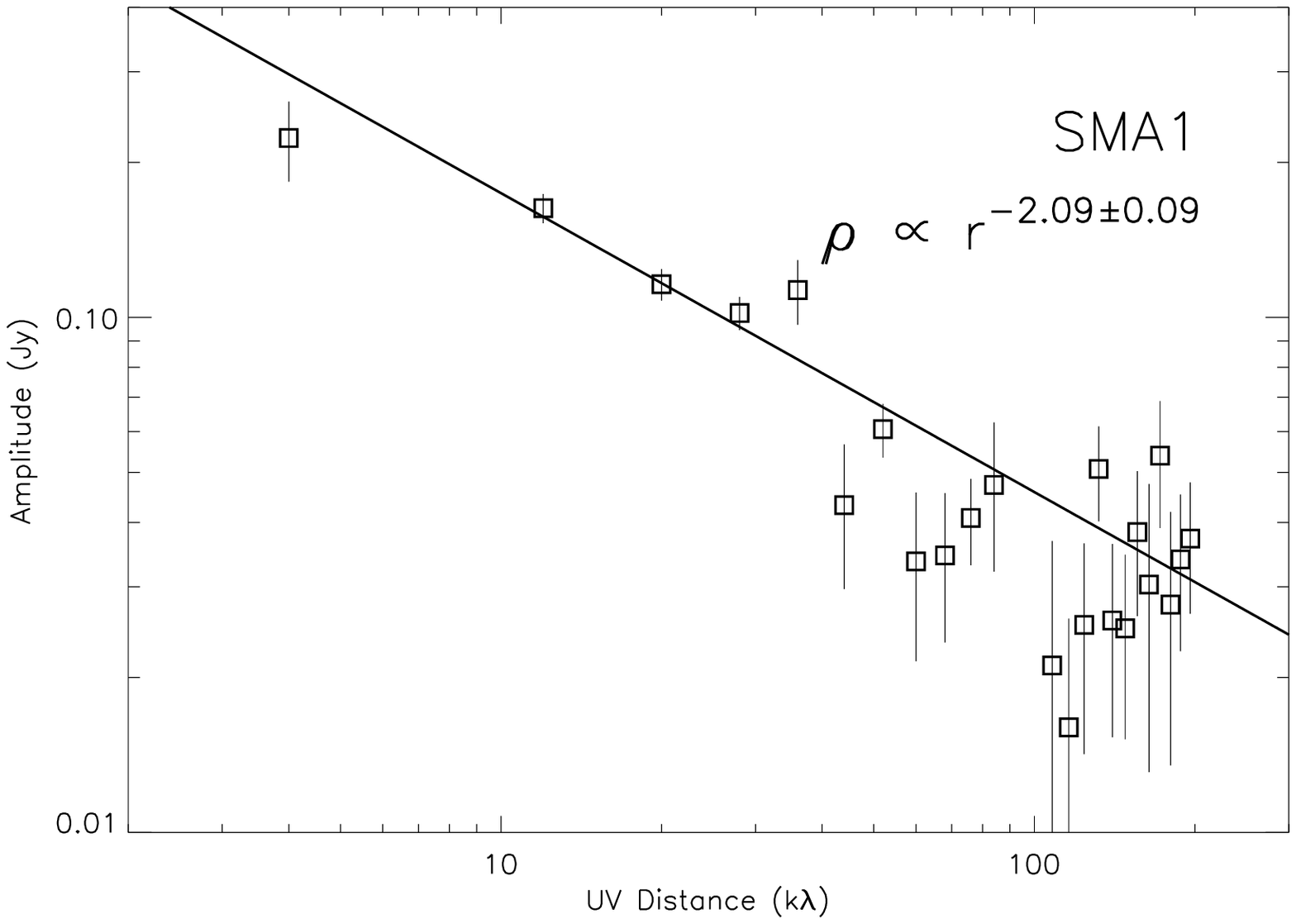}
\includegraphics[width=0.5\textwidth,angle=0]{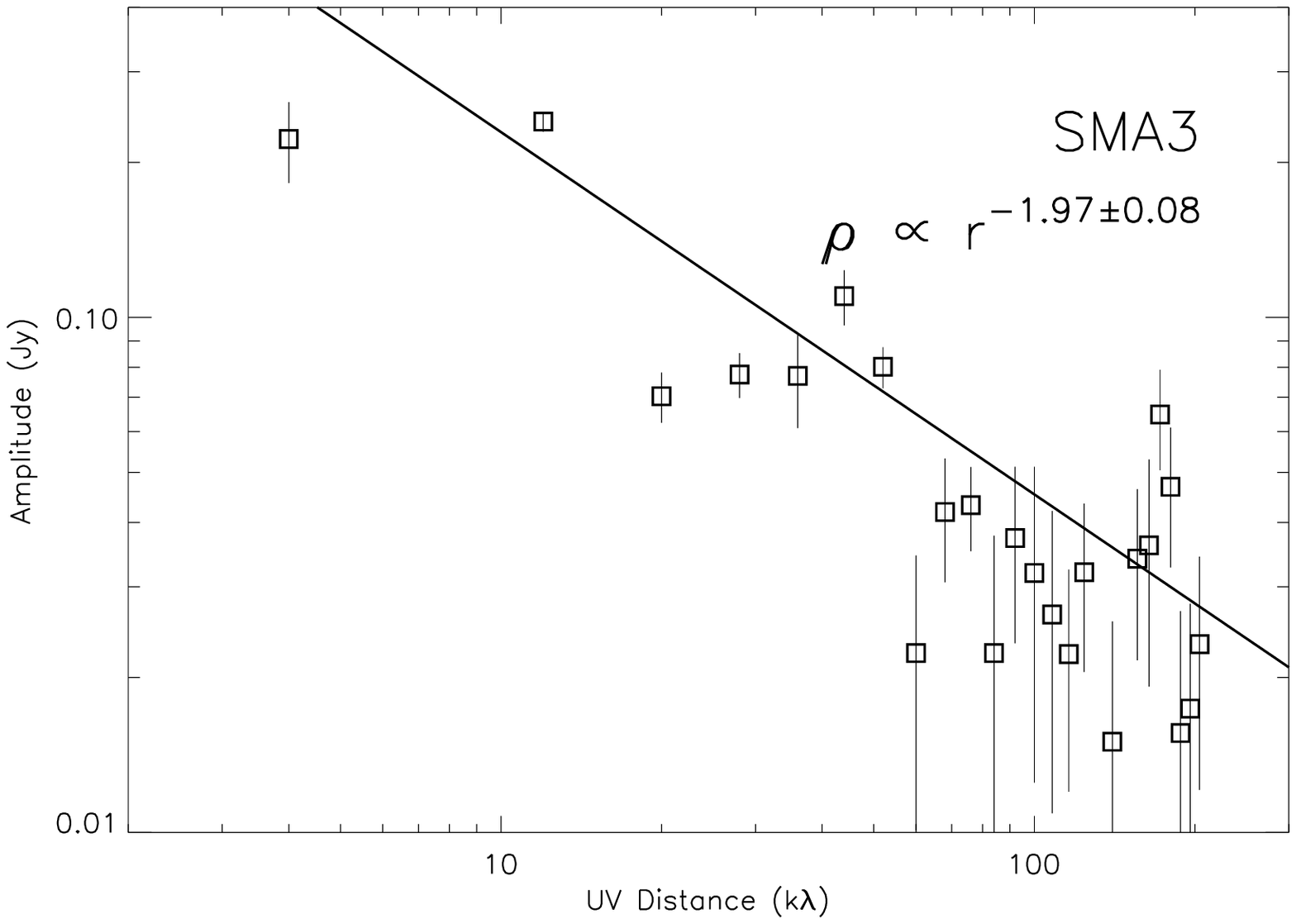}
\includegraphics[width=0.5\textwidth,angle=0]{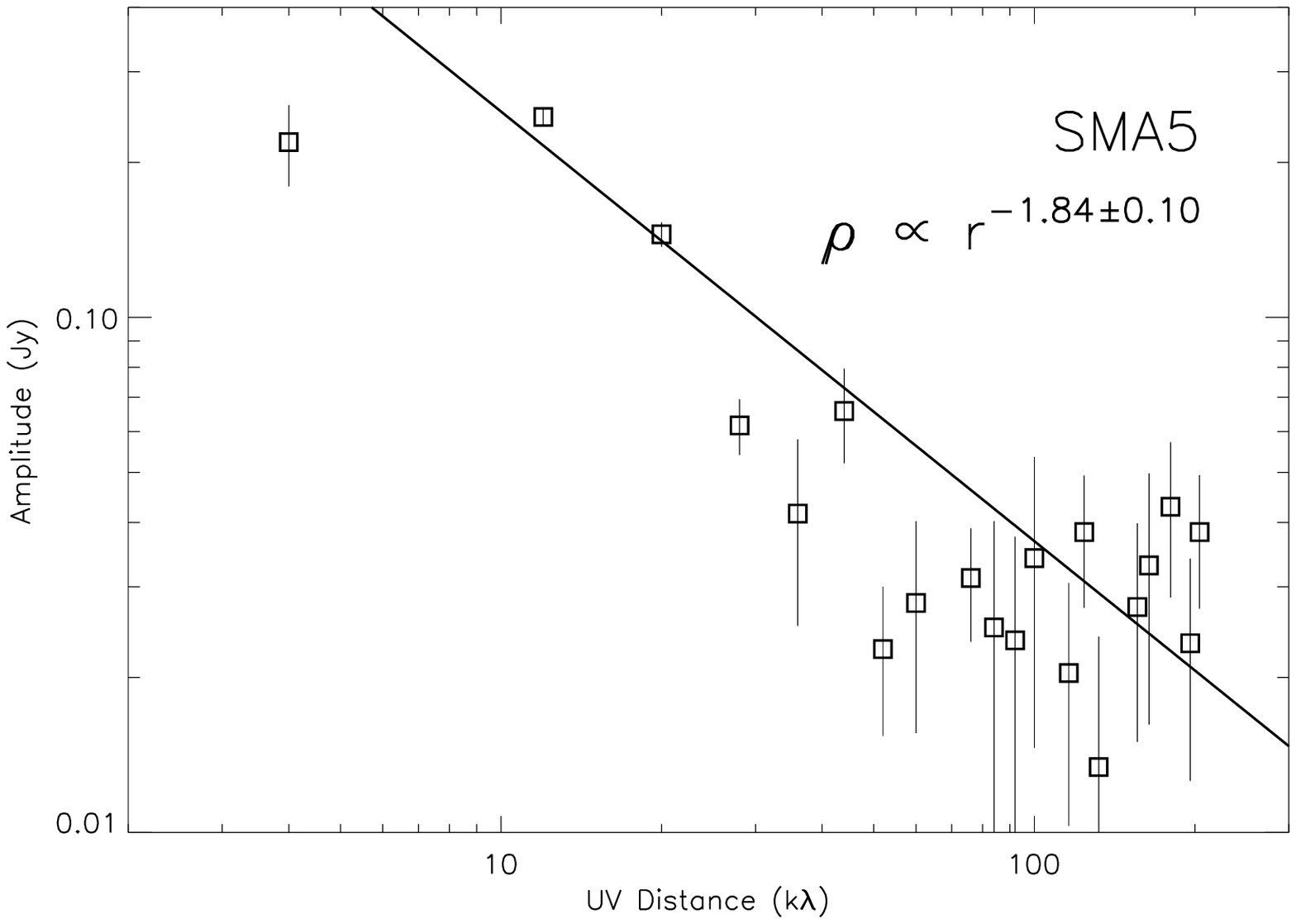}
\caption{Amplitude distribution against \uv\ distance of cores SMA1, SMA3, and SMA5.
\label{fig:uvamp}}
\end{figure}

Fig.~\ref{fig:uvamp} plots the amplitude versus \uv\ distance for SMA1, SMA3, and SMA5.
The amplitude is a vector average of the visibility data
in a concentric annuli defined by a bin of 8\,k$\lambda$.
Least-squares fitting yields
$b=2.09 \pm 0.09$ for SMA1, $1.97 \pm 0.08$ for SMA3, and $1.84 \pm 0.10$ for SMA5.
The results are consistent with \cite{zhang09} measured from the 230 GHz continuum data,
where they derived $b=2.1 \pm 0.2$ for SMA4.
The density profiles are similar to that of an isothermal sphere in hydrostatic equilibrium,
where the radial density scales as $\propto r^{-2}$.
We note that
the large error bars and discrepancy at long \uv\ distances reflect sensitivity limits
and may be also partly due to unresolved weak sources in the vicinity.

\subsection{Spectral line data and CO outflows \label{sect:outflow}}
\subsubsection{Morphology \label{subsect:outflow_morphology}}

\begin{figure}
\includegraphics[height=\textwidth,angle=-90]{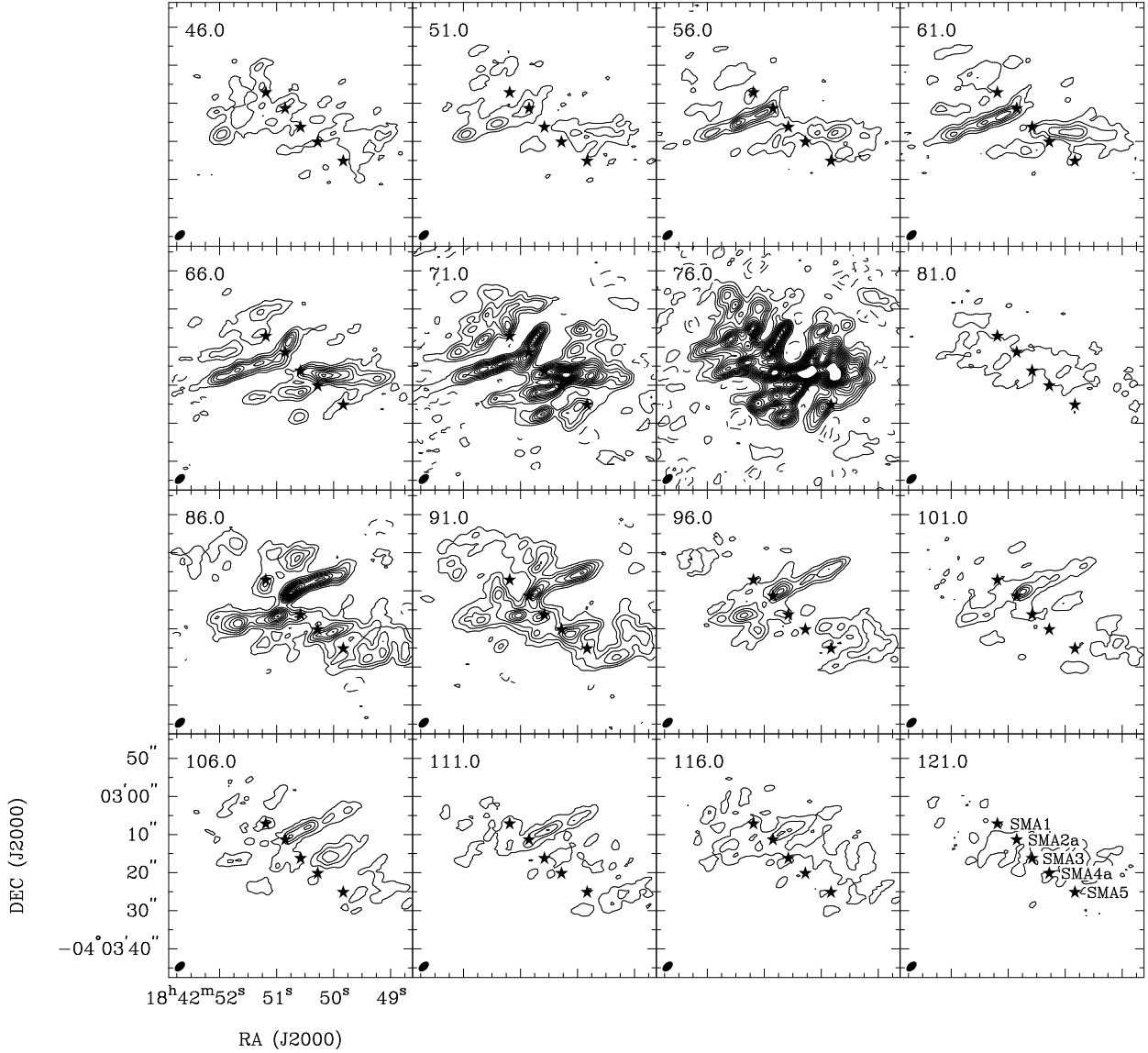}
\caption{CO\,(3--2) channel maps.
The image is made by averaging the emission over a 5\,\kms-1 velocity interval,
with the central velocity labeled on each panel.
The contours start at 5$\sigma$ (240\,mJy) and 
step by $\pm 5\sigma$.
The filled ellipse on the bottom left corner represents the synthesized beam.
The stars mark the presumable outflow driving sources as labeled on the last panel (see the text).
\label{fig:chmap}}
\end{figure}

\begin{figure}
\includegraphics[width=\textwidth,angle=0]{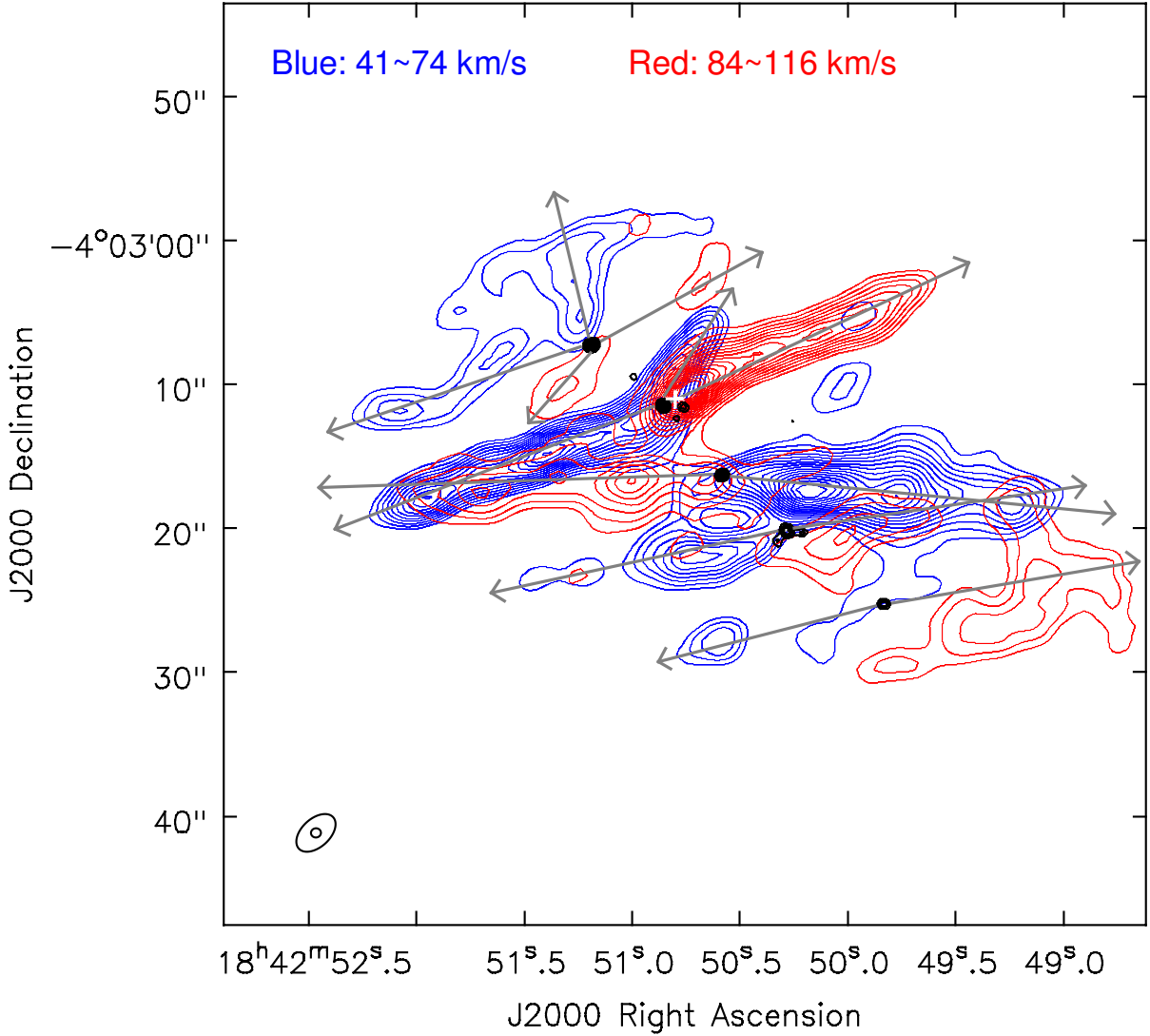}
\caption{
Integrated CO\,(3--2) emission overlaid on the 0.88\,mm continuum emission
of the G28.34-P1 region.
The velocity ranges used for integration are [41, 74]\,\kms-1 for the blueshifted lobe (blue contours),
and [84, 116]\,\kms-1 for the redshifted lobe (red contours).
The CO contours are $\pm(2, 2.5, 3, ..., 9.5) \times 10\%$ of the blue lobe peak,
39\,Jy\,beam$^{-1} \cdot$\kms-1.
The continuum contours start at 4\,mJy ($5\sigma$) and increase by a step of
1.6\,mJy ($2\sigma$).
The white cross represents the H$_2$O maser detected by VLA \citep{wang06}.
[Follow up observations have resolved this maser spot into two,
spatially coincident with SMA2a and SMA2b,
respectively (Wang et al. 2011 in prep.)]
The arrows sketch outflow directions.
The synthesized beams indicated in the lower left corner are
$3''.23\times 1''.96$, $\rm{PA}=-47^{\circ}.4$ for the CO image,
while $0''.69\times 0''.64$, $\rm{PA}=-83^{\circ}.1$ for the continuum image.
\label{fig:outflows}
}
\end{figure}

Among the entire 8\,GHz SMA band,
only CO\,(3--2) is detected above $3\sigma$
($1\sigma \approx 0.1$\,K at 1 \kms-1 resolution).
Nevertheless, CO\,(3--2) reveals high-velocity outflowing gas in P1.
Fig.~\ref{fig:chmap} presents the CO\,(3--2) channel maps, and Fig.~\ref{fig:outflows} plots
the integrated blueshifted emission in blue contours,
and redshifted emission in red contours, 
superposed on the continuum emission.
One can find in Fig.~\ref{fig:chmap} that
the extended CO emission close to the systemic velocity (78.4 \kms-1) is filtered out.
This effect actually helped us identify outflows at high velocities.
It is intriguing that all the five dust cores are associated with CO outflows,
majority of which are bipolar, jet-like outflows.

The outflows are centered on the dust cores and are generally oriented cross the major axis of the filament.
The most prominent outflow is centered on core SMA2, 
with a blue lobe emanating toward southeast (SE; see channels of 51--71 \kms-1 in Fig.~\ref{fig:chmap}),
and a red lobe toward northwest (NW; 86--116 \kms-1).
In addition, a minor blue lobe is also seen at NW (66--76 \kms-1).
The outflow centered on SMA3 is in an east-west orientation,
with its blue lobe emanating westbound (51--71 \kms-1)
and its red lobe eastbound (86--96 \kms-1).
SMA4 lies in the center of a NW-SE outflow,
with a SE blue lobe (66--76 \kms-1)
and a NW red lobe (86--91 \kms-1).
SMA1 is associated with a quadrupolar outflow.
Its two blue lobes are seen at 66--71 \kms-1,
emanating toward SE and north, respectively.
Its two red lobes are seen from 86 through 91 \kms-1 toward NW and south,
respectively.
The outflow centered on SMA5 has an orientation almost parallel to the SMA4 outflow.
This outflow is weak compared to others,
and its SE blue lobe and NW red lobe can be barely seen at
66--71 \kms-1 and 86--91 \kms-1, respectively.
Orientation of all the outflows are sketched as arrows in Fig.~\ref{fig:outflows}.

The driving sources of these outflows can be identified geometrically.
Cores
SMA1 to SMA5 are roughly located at the geometric centers of the outflows, respectively.
While SMA1, SMA3 and SMA5 are unresolved, SMA2 and SMA4 are resolved into three condensations.
In these two cases, we assign the strongest and/or central condensation
as the driving source of the relevant outflow.
Hence, the outflows are likely driven by 
condensations
SMA1, SMA2a, SMA3, SMA4a, and SMA5.

The CO outflows shown in Figs. \ref{fig:chmap} and \ref{fig:outflows} 
consist of pairs of knots which are geometrically
symmetric with respect to the central protostar.
Illustrated in Fig.~\ref{fig:outflows},
outflows SMA2a and SMA3
have four pairs of knots, and outflow SMA4a has two pairs.
The knots are equally spaced by about 0.16\,pc.
These knots in outflows may represent outbursts due to disk variability arising
from episodic, unsteady accretion (e.g., HH 211, \citealt{lee10}; HH 80-81, \citealt{qiu2009}).
With the spacing and maximum outflow velocity $v_{\rm max}$,
we estimate the period between two outbursts to be about 4000\,yr.
This value lies in the range of roughly $10^3-10^4$ yr used in
episodic accretion models for low mass stars \citep{Baraffe2010,Zhu2010}.

The outflows show slight bending 
(Fig.~\ref{fig:outflows}).
In the outflow SMA3,
the red lobe bends  $5^{\circ}$ toward the south,
and the blue lobe bends  $6^{\circ}$ toward the south,
making the outflow \textsf{C}-shaped.
In outflows SMA2a and SMA4a, and probably SMA1,
the knots seem to trace the propagation of a periodically wiggled,
\textsf{S}-shaped molecular jet,
most evident at the closest knotty pairs near the protostar.
Part of this effect may be contaminated by multiple outflows (e.g., in outflow SMA2a).
Nevertheless, the
mirror symmetric (\textsf{C}-shaped) wiggles and
point symmetric (\textsf{S}-shaped) wiggles
may indicate different bending mechanisms like
jet precession, orbital motion of the jet source,
and Lorentz forces \citep{Fendt98, Masciadri02, Raga09}.
Similar bending effects have been 
observed in a number of Herbig-Haro jets \citep[e.g., HH 211, ][]{lee10}
and some of the most collimated high-mass molecular jets \citep[e.g.,][]{Su2007}.
Because the wiggles are not prominent in CO(3--2),
we do not discuss this effect further in this paper.

\subsubsection{Physical properties \label{subsect:outflow_para}}

We calculate the physical parameters of each outflow,
including mass, momentum, energy, as well as dynamical age and outflow rate.
Assuming local thermodynamic equilibrium and optically thin CO emission in the line wings,
we first derive the CO column density following \cite{Garden91},
$$
N_{\rm CO} ({\rm cm^{-2}}) = 4.81 \times 10^{12}\, (T_{\rm ex}+0.92)\,
{\mathrm {exp}}\left(\frac{33.12}{T_{\rm ex}}\right)\int
T_{\rm B}\,dv\,\,,
$$
where $dv$ is the velocity interval in \kms-1,
and $T_{\rm ex}$ and $T_{\rm B}$ are excitation temperature and brightness temperature in K, respectively.
We take the gas temperatures to be the excitation temperatures,
{i.e.}, 16\,K for the SMA2a outflow
and 13\,K for other outflows.
The outflow mass, momentum, energy, dynamical age, and outflow rate are then given by
$$
M = d^2\,[\frac{\rm H_2}{\rm CO}]\,\overline{m}_{\rm H_2}
\int_{\Omega}N_{\rm CO}(\Omega^\prime)d\Omega^\prime\,,
$$
$$
P = Mv\,,
$$
$$
E = \frac{1}{2}Mv^2\,,
$$
$$
t_{\rm dyn}=\frac{L_{\rm flow}}{v_{\rm max}},
$$
$$
\dot{M}_{\rm out} = \frac{M}{t_{\rm dyn}},
$$
where $d$ is the source distance,
$\overline{m}_{\rm H_2}$ is the mean mass per hydrogen molecule assumed to be 2.33 atomic units,
$\Omega$ is the total solid angle that the flow subtends,
$v$ is the outflow velocity with respect to the systemic velocity of 78.4\,\kms-1,
$v_{\rm max}$ is the maximum outflow velocity,
and $L_{\rm flow}$ is the flow length.
We adopt an empirical $[\frac{\rm H_2}{\rm CO}]$ abundance ratio of $10^4$ \citep{blake87}.

Table \ref{tab:outflows} lists derived physical parameters for each outflow.
These outflows show symmetry in blue- and redshifted lobes.
The energetics are small compared to outflows observed in high mass protostellar
objects \citep{Beuther02c,zhang2005},
but are similar to outflows emanating from
nearby ($\lesssim 1$\,kpc) intermediate-mass hot cores
(IMHCs: IRAS 22198+6336, \citealt{sanchez10};
NGC 7129-FIRS 2, \citealt{Fuente05};
IC 1396 N, \citealt{Neri07}).
Particularly, the outflow rate \Mout, which amounts to $(4-47)\times 10^{-6}$\,\msun\,yr$^{-1}$,
and the flow momentum $P$, which ranges from 1 to 9 \msun\,\kms-1,
are of the same order to all the three IMHC outflows known to date.
The dynamic timescales are $(1.5-3.4)\times 10^4$ yr for all the outflows.
These values are comparable to massive outflows but are one order of magnitude higher than the IMHC outflows.
Because observations are limited by noise and gas at higher velocities are possible,
\tdyn\ is a lower limit of the outflow age and the accretion history of the dominant protostar.
Theoretical models suggest that the massive star formation process takes a few times of the
free-fall timescale
(\tff$=\sqrt{\frac{3\pi}{32G\rho}}=\frac{3.66\times 10^7 \rm{yr}}{\sqrt{n(\rm{cm}^{-3})}}$)
of the core.
\cite{Beuther02c} observed a rough equality between \tdyn\ and \tff.
For the P1 cores, we obtain \tff\ of $(1.2-6.7)\times 10^4$
yr based on the densities given in Table \ref{tab:frags}.
We see that \tff\ approximates \tdyn\ well,
and this supports the idea that flow ages are good estimates of protostar lifetimes \citep{Beuther02c}.

The SMA2a, SMA3, and SMA4a outflows are so well collimated that they resemble ``molecular jets''
commonly found in low-mass star-forming regions
(e.g., L1157, \citealt{zhang2000};
OMC-1S, \citealt{zapata05};
HH 211, \citealt{lee07}).
Recently, interferometric observations 
have revealed collimated outflows
in high-mass star-forming regions
(e.g., AFGL 5142, \citealt{Hunter99,zhang2007};
IRAS 20126+4104, \citealt{Cesaroni99, Shepherd2000};
IRAS 05358+3543, 19410+2336, 19217+1651, \citealt{Beuther02a, Beuther03, Beuther04};
HH80-81, \citealt{qiu2009}).
To measure the collimation,
we take the ratio of outflow total length (blue + red lobes) over width measured in
Fig.~\ref{fig:outflows} as collimation factor.
We obtain 14, 10, and 7 for SMA2a, SMA3, and SMA4a outflows,
over scales of 0.5--1\,pc.
These values are comparable to that of the most collimated low-mass outflows
and are higher than that of the high-mass outflows 
discovered by single dishes and previous interferometers 
\citep{Wu04, Beuther02c, qiu2009}.
\cite{Beuther07} reported one outflow in IRAS 05358+3543
with a collimation factor of $\sim 10$
and claimed that it was the first massive outflow
observed with such a high degree of collimation at a scale of 1\,pc.
In G28.34-P1, the SMA2a and SMA3 outflows
are among the best collimated molecular outflows yet discovered
in high-mass star-forming regions over parsec scales.
We note that these collimation factors should be considered as lower limits because
the outflow lengths and widths are not corrected for inclination angle, 
and are not deconvolved with synthesized beam.

\subsection{Massive star formation through an intermediate-mass stage \label{sect:discuss}}

The presence of highly collimated outflows demonstrates a protostellar nature of the 
condensations in G28.34-P1.
\cite{wang08} reported a luminosity of $10^2$\,\lsun\ for the 24\,\um\ source associated with SMA2.
\cite{rath10} obtained a luminosity of $2.1 \times 10^3$\,\lsun\ for the entire P1 clump.
These luminosities set a limit on the mass of the embedded stars of 3--7 \msun, if
they are at the zero-age main sequence.
The stellar masses put these objects at an intermediate-mass stage.
The presence of highly collimated outflows also favors a disk-mediated accretion scenario of
massive mass star formation \citep{zhang2005, cesaroni2007} in a manner similar to
low-mass star formation.
In addition, the high collimation also indicates that these outflows are at very early stages
since the opening angle of outflows widens as they evolve.

The G28.34-P1 condensations are comparable in mass to the nearby intermediate-mass ``cores''
(2--5\,\msun) at a similar spatial extent ($2300 \times 1600$\,AU; \citealt{sanchez10}).
The non-detection of organic molecules (CH$_3$OH, CH$_3$CN, etc.) in P1 is likely due to 
the sensitivity limit of our observations.
\cite{sanchez10} observed copious hot core molecular line emissions toward I22198,
an intermediate-mass star with a luminosity of 370 \lsun\
at a distance of 0.76\,kpc. The typical brightness temperature of the lines is 0.8\,K.
If we scale I22198 to the distance of G28.34, the emission would become 40 times fainter,
well below our detection limit of 0.1 K per 1 \kms-1\ resolution \citep[also true for ][]{zhang09}.
Therefore, it is possible that G28.34-P1 may have already experienced an increase
in abundance of organic molecules as expected from protostellar heating, similar to I22198.
Deep spectroscopy with ALMA will be able to verify the presence of hot core type
molecules in G28.34-P1.

Despite the similarities between the G28.34-P1 condensations and the intermediate-mass
star forming ``cores'', there is an important distinction between the two: 
The G28.34-P1 condensations are embedded in compact ($\sim 0.1$\,pc),
dense cores ($\sim 10^7$\,\cm3, see Table \ref{tab:frags}) 
with several tens of solar masses,
which are hierarchically embedded in a massive, 
dense ($3 \times 10^5$\,\cm3, Table \ref{tab:frags}) molecular clump of over $10^3$ \msun, 
an order of magnitude larger than the mass in a typical intermediate-mass star-forming region. 
With a typical star formation efficiency of 30\% in dense gas
and a standard stellar initial mass function,
it is natural to expected that massive stars will eventually emerge in
some of the condensations (if not all), at the end of protostellar accretion.
At that time, P1 would become a massive star cluster.
In the context of evolution, an immediate comparison to G28.34-P1 is
the northern clump G28.34-P2 in the same cloud complex, which
may represent the future of P1. 
With a similar clump mass, 
P2 has a higher gas temperature of 30 K,
and a higher luminosity of $2.6 \times 10^4$ \lsun\ \citep{rath10}, 
equivalent to a zero-age main-sequence mass of 14 \msun.
The SMA observations detected molecular lines from CH$_3$OH and
CH$_3$CN in the P2 cores, while none in the P1 cores \citep{zhang09}.
Likewise, other massive molecular clumps (e.g., W33A, \citealt{galvan2010}; G31.41, \citealt{girart2009};
NGC 6334, \citealt{hunter2006}; G10.6, \citealt{liu2010}) that have luminosities of $10^4 - 10^5$ \lsun,
and are associated with radio continuum emission, may represent an even more evolved phase of
massive star formation at which high mass protostellar objects have already emerged.

The large mass reservoir (22--64\,\msun\ at $\sim0.1$\,pc; 1000\,\msun\ at $<1$\,pc) in G28.34-P1
and the outflows suggest that the condensations are, and will remain, 
actively accreting materials through the ambient medium.
Materials in the cores fall onto the embedded protostars through disks
and at the same time materials in the clump fall onto the cores.
Observations of G28.34-P1 seem to indicate a continuous accretion
over a spatial scale that spreads two orders of magnitude, from 1 to 0.01\,pc,
and finally reaches onto the protostars.
This is a relatively new picture,
but has also been reported in another IRDC,
G30.88+0.13 \citep{zhang2011}.
A comparison with massive and more luminous molecular clumps indicates that
the condensations will grow in mass, and likely emerge as massive stars eventually.
The embedded protostars appear to be still intermediate-mass objects and 
are on their way to eventual massive stars.
Our observations of the G28.34-P1 clump provide perhaps the first glimpse
of an intermediate-mass stage of massive, clustered star formation.

\section{Conclusion}
We present an observational study of the P1 clump in IRDC G28.34+0.06
using the SMA at the 345\,GHz band.
Our main findings are summarized as follows.

1. High resolution and high sensitivity image at 0.88 mm reveals
hierarchical fragmentation at 1\,pc and 0.1\,pc scales. 
The mass of the dust cores and condensations 
are much larger than the 
relevant thermal Jean masses.
The clump fragmentation is consistent with a cylindrical collapse.

2. All five dust cores are driving highly collimated CO\,(3--2) outflows,
indicating a protostellar nature of the condensations embedded in the cores.
Bipolar, jet-like outflows suggest a disk-mediated accretion of high-mass star formation.

3. The relatively low luminosity and a lack of detection of organic molecules indicate that 
the condensations in P1 underline an intermediate-mass stage of massive star formation
and will likely form massive stars eventually.

\acknowledgements
We appreciate an anonymous referee's valuable comments that helped clarify the manuscript.
K.W. acknowledges the support from the SMA predoctoral fellowship and the China Scholarship Council.
Q.Z. acknowledges the support from the Smithsonian Institution Endowment Funds.
This research is funded in part by NSFC grants 10873019 and 11073003.

{\it Facility:} \facility{SubMillimeter Array}

\begin{deluxetable}{lcccllcc}
\tablecolumns{8}
\tablewidth{0pc}
\tabletypesize{\footnotesize}
\tablecaption{Summary of the SMA Observations \label{tab:obs}}
\tablehead{\colhead{Array} & \colhead{UT date} & \colhead{Antenna} &
\colhead{Bandwidth} & \colhead{Bandpass} & \colhead{Flux} & \colhead{$T_{\rm sys}$} & \colhead{$\tau _{\rm 225\,GHz}$ \tablenotemark{a}} \\
\colhead{} & \colhead{(yyyymm)} & \colhead{No.} &
\colhead{(GHz)} & \colhead{Calibrator} & \colhead{Calibrator} & \colhead{(K)} & \colhead{} }
\startdata
COM    &20090426 &7 &2     &3C273    &Ceres                &200-400 &0.09-0.15 \\
COM    &20090503 &7 &2     &3C273    &Ceres                &160-280 &0.05 \\
EXT    &20090806 &8 &4     &3C454.3  &MWC349               &260-500 &0.09-0.14 \\
EXT    &20090826 &7 &4     &3C454.3  &Uranus               &120-250 &0.06-0.07  \\
SUB    &20090905 &5 &4     &3C454.3  &Ganymede, Callisto    &160-300 &0.05-0.07 \\
SUB    &20090923 &7 &4     &3C454.3  &Ganymede, Callisto    &150-300 &0.07-0.09 \\
\enddata
\tablenotetext{a}{Zenith opacity measured from water vapor monitors mounted on CSO or JCMT.}
\end{deluxetable}

\begin{deluxetable}{lccccrrrcc}
\tablecolumns{10}
\tablewidth{0pc}
\tabletypesize{\footnotesize}
\tablecaption{Properties of the Condensations \label{tab:cores}} 
\tablehead{\colhead{Name} & \colhead{RA(J2000)} & \colhead{Dec(J2000)} &
\multicolumn{2}{c}{Size} & \colhead{P.A.} & \colhead{$F_{\rm peak}$} & \colhead{$F_{\rm int}$} & 
\multicolumn{2}{c}{Mass/\msun} \\
\colhead{} & \colhead{h:m:s} & \colhead{d:m:s} &
\multicolumn{2}{c}{a$'' \times \rm{b}''$} & \colhead{$^{\circ}$} & \colhead{mJy} & \colhead{mJy}
& \colhead{$\beta=1.5$} & \colhead{$\beta=2$}}
\startdata
SMA1	&	18:42:51.19	&	-04:03:07.2	&	0.82	&	0.68	&	-63.6	&	25.3	&	31.6	&	10.6	&	19.8	\\
SMA2a	&	18:42:50.85	&	-04:03:11.4	&	0.74	&	0.70	&	55.2	&	15.7	&	18.3	&	4.3	&	8.1	\\
SMA2b	&	18:42:50.76	&	-04:03:11.5	&	1.18	&	0.68	&	56.8	&	8.5	&	15.5	&	3.6	&	6.8	\\
SMA2c	&	18:42:50.79	&	-04:03:12.4	&	0.81	&	0.62	&	60.5	&	5.1	&	5.8	&	1.4	&	2.6	\\
SMA2d	&	18:42:50.99	&	-04:03:09.5	&	0.94	&	0.63	&	62.2	&	5.1	&	6.7	&	1.6	&	3.0	\\
SMA3	&	18:42:50.58	&	-04:03:16.3	&	0.76	&	0.63	&	-82.3	&	14.4	&	15.6	&	5.2	&	9.8	\\
SMA4a	&	18:42:50.28	&	-04:03:20.2	&	0.78	&	0.75	&	38.1	&	12.2	&	16.0	&	5.3	&	10.0	\\
SMA4b	&	18:42:50.32	&	-04:03:21.0	&	1.30	&	0.75	&	33.2	&	6.8	&	15.0	&	5.0	&	9.4	\\
SMA4c	&	18:42:50.20	&	-04:03:20.3	&	0.93	&	0.55	&	-84.7	&	7.4	&	8.5	&	2.8	&	5.3	\\
SMA5	&	18:42:49.83	&	-04:03:25.2	&	0.91	&	0.74	&	78.8	&	16.6	&	25.1	&	8.4	&	15.7	\\
\enddata
\end{deluxetable}

\begin{deluxetable}{lcc ccc}
\tablecolumns{6}
\tablewidth{0pc}
\tabletypesize{\small}
\tablecaption{Clump/Core Fragmentation \label{tab:frags}}
\tablehead{\colhead{Source} & \colhead{$T$ \tablenotemark{a}} & \colhead{$n$(H$_2$)} &
\colhead{$M_J$} & \colhead{$R_J$} & \colhead{Fragment Mass \tablenotemark{b}}\\
\colhead{} & \colhead{K} & \colhead{cm$^{-3}$} &
\colhead{\msun} & \colhead{pc} & \colhead{\msun} }
\startdata
P1  & 13  & $3.0\times10^5$  & 0.8  &0.04  & 22--64\\
P1-SMA2  & 16  & $9.6\times10^6$  & 0.2 &0.008  & $>$1.4--4.3\\
\enddata
\tablenotetext{a}{Gas temperatures are derived from VLA \nh3 (1,1) and (2,2) lines \citep{wang08}.}
\tablenotetext{b}{Observed core/condensation masses are calculated using a dust opacity index of
$\beta = 1.5$.}
\end{deluxetable}

\begin{deluxetable}{l cc cc cc cc cc cc}
\tabletypesize{\scriptsize}
\tablecolumns{11}
\tablewidth{0pt}
\tablecaption{Derived Outflow Parameters \label{tab:outflows}}
\tablehead{\colhead{Parameter \tablenotemark{a}}
& \multicolumn{2}{c}{SMA1}
& \multicolumn{2}{c}{SMA2a}
& \multicolumn{2}{c}{SMA3}
& \multicolumn{2}{c}{SMA4a}
& \multicolumn{2}{c}{SMA5} \\
& Blue & Red
& Blue & Red
& Blue & Red
& Blue & Red
& Blue & Red}
\startdata
$v$ (\kms-1) \tablenotemark{b}        &[67,76] &[85,93] &[41,76] &[84,116] &[50,71] &[85,99] &[64,72] &[85,107] &[66,73] &[86,93] \\
$M$ (\msun)             	      &0.20   &0.07   &0.38    &0.23    &0.22    &0.18    &0.06   &0.12   &0.05   &0.06 \\
$P$ (\msun\,\kms-1)                   &1.12   &0.71   &5.09    &4.04    &3.17    &2.05    &0.52   &1.50   &0.35   &0.63 \\
$E$ (\msun\,km$^2$ s$^{-2}$)          &3.85   &3.62   &53.46   &46.18   &26.96   &13.06   &2.57   &11.15  &1.40   &3.40 \\
$L_{\rm flow}$ (pc) \tablenotemark{c} &0.40   &0.27   &0.51    &0.48    &0.56    &0.51    &0.44   &0.43   &0.34   &0.39 \\
$t_{\rm dyn}$ ($10^4$\,yr)            &3.43   &1.81   &1.33    &1.25    &1.93    &2.42    &2.99   &1.47   &2.68   &2.61 \\
\Mout ($10^{-5}$\,\msun\,yr$^{-1}$)   &0.58   &0.41   &2.86    &1.84    &1.14    &0.74    &0.20   &0.82   &0.19   &0.23 \\
\enddata
\tablenotetext{a}{Parameters are not corrected for inclination angle.}
\tablenotetext{b}{Range of velocities used to derive outflow parameters.}
\tablenotetext{c}{Projected length of the longer lobe in case of two lobes.}
\end{deluxetable}

\end{document}